\documentclass{article}
\usepackage{spconf,amsmath,graphicx}
\usepackage{xcolor}
\usepackage{colortbl}
\usepackage{multirow}
\usepackage{booktabs}

\usepackage[numbers,sort&compress]{natbib}
\usepackage[textsize=tiny,textwidth=0.7in, disable]{todonotes}

\title{
TidyVoice: A Curated Multilingual Dataset for Speaker Verification Derived from Common Voice}
\name{Aref Farhadipour$^{1}$, Jan Marquenie$^{2}$, Srikanth Madikeri$^{1}$, and Eleanor Chodroff$^{1}$\thanks{This research was supported by funding from SNF Grant PR00P1\_208460 to EC.}}
\address{$^{1}$Department of Computational Linguistics, University of Zurich \\ $^{2}$Mobile Dialog Systems, Otto-von-Guericke-University Magdeburg \\
\{aref.farhadipour, srikanth.madikeriraghunathan, eleanor.chodroff\}@uzh.ch, jan.marquenie@ovgu.de}

\begin{document}
%
\maketitle
\begin{abstract}
The development of robust, multilingual speaker recognition systems is hindered by a lack of large-scale, publicly available and multilingual datasets, particularly for the read-speech style crucial for applications like anti-spoofing. 
To address this gap, we introduce the TidyVoice dataset derived from the Mozilla Common Voice corpus after mitigating its inherent speaker heterogeneity within the provided client IDs. TidyVoice currently contains training and test data from over 212,000 monolingual speakers (Tidy-M) and around 4,500 multilingual speakers (Tidy-X) from which we derive two distinct conditions. The Tidy-M condition contains target and non-target trials from monolingual speakers across 81 languages. The Tidy-X condition contains target and non-target trials from multilingual speakers in both same- and cross-language trials. 
We employ two architectures of ResNet models, achieving a 0.35\% EER by fine-tuning on our comprehensive Tidy-M partition. Moreover, we show that this fine-tuning enhances the model's generalization, improving performance on unseen conversational interview data from the CANDOR corpus. 
The complete dataset, evaluation trials, and our models are publicly released to provide a new resource for the community.
\end{abstract}
\begin{keywords}
Speaker verification, benchmark, multilingual, cross-language
\end{keywords}
\section{Introduction}
\label{sec:intro}

Speaker recognition has become an integral technology in myriad applications, ranging from biometric security and access control to speaker diarization and personalization of smart devices. 
The advent of deep learning has led to state-of-the-art systems in this field \cite{far2024analys,desplanques20_interspeech,barahona2025analysis,lin2024voxblink2}, achieving impressive performance on established benchmarks \cite{nagrani2020voxceleb, chung2018voxceleb2, lee20232022}. 
A significant limitation of many current systems is their reliance on datasets that are predominantly monolingual, with a heavy bias towards English~\cite{hutiri2022bias}. 
Moreover, while in-the-wild benchmarks are excellent for capturing spontaneous speech, critical applications such as anti-spoofing, where systems are often evaluated using text-prompted utterances, also require training and test data that reflects a read-speech style. 
This dual challenge, consisting of the demand for broad linguistic diversity and the need for a large-scale, read-speech corpus, poses a major obstacle to deploying robust and fair speaker recognition systems globally.



\begin{figure}[t]  
  \centering
  \newcommand{\figscale}{0.40}

  \includegraphics[scale=\figscale,trim=35 10 35 35,clip]{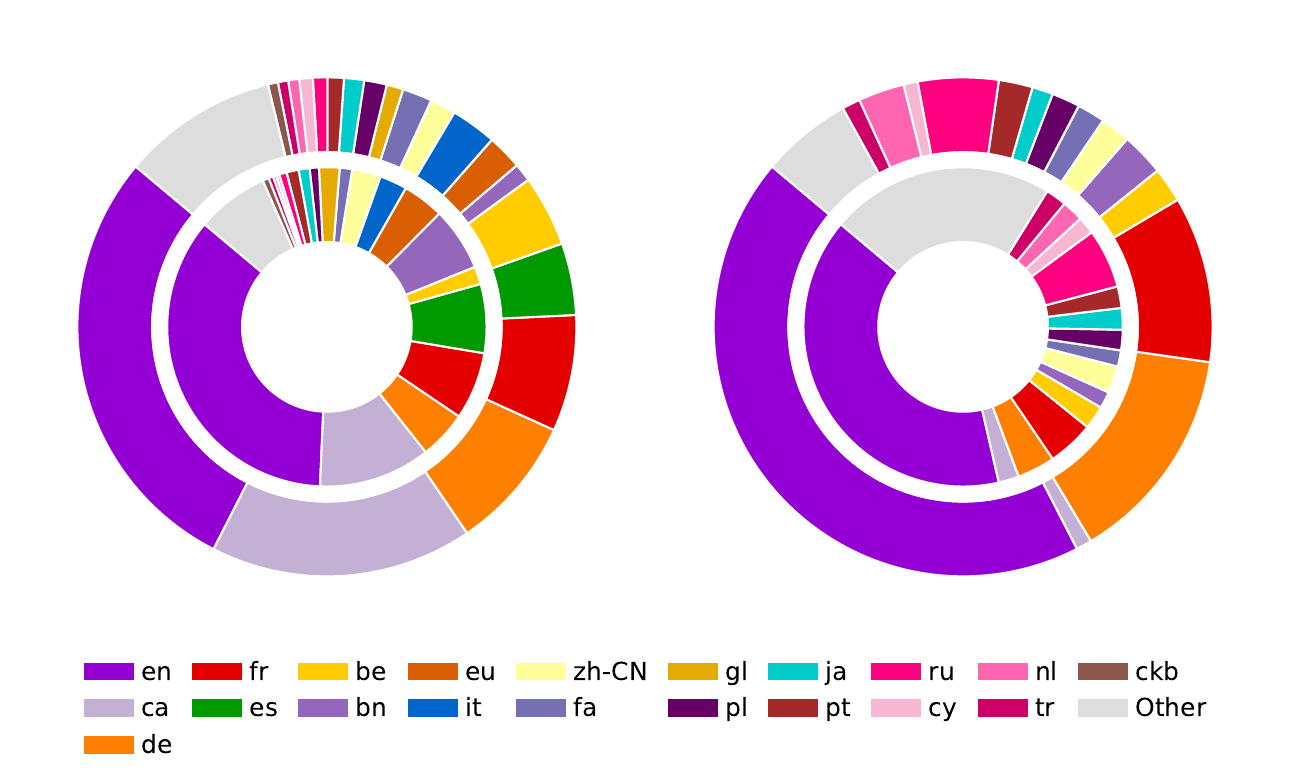}%

  \caption{Language distribution in the train (outer circle) and test (inner circle) sets for the Tidy-M (right) and Tidy-X (left) partitions. For example, the training data for Tidy-M includes around 40K and 940 speakers for ``en'' and ``tr'' respectively.}
  \label{fig:diagrams}
\end{figure}
In recent years, two main families of benchmarks have been instrumental in advancing the state of the art for multilingual speaker recognition and verification: the in-the-wild VoxCeleb series and the domain-specific NIST SRE corpora. 
Table~\ref{tab:dataset_comparison} summarizes these key benchmarks and contrasts them with our contributions. 
The VoxCeleb~\cite{nagrani2017voxceleb, chung2018voxceleb2} and the more recent VoxBlink~\cite{lin2024voxblink1, lin2024voxblink2} series have become de-facto standards for benchmarking. 
Sourced from YouTube videos, these datasets pushed the boundaries of scale, culminating in VoxBlink2 with over 111,000 speakers. While highly multilingual by nature of their collection, their primary characteristic is diverse and spontaneous in-the-wild speech. 


In parallel, evaluations driven by the NIST Speaker Recognition Evaluation (SRE) series have been crucial. The collective training data, referred to as the NIST SRE Superset~\cite{sadjadi2021nist}, comprises around 7,000 speakers from a wide variety of languages, with conditions targeting conversational telephone speech (CTS) and audio from video (AfV) domains. 
While serving as a critical resource, the SRE Superset and SRE evaluation sets are licensed through the LDC and require a financial investment.

In the present work, we introduce the TidyVoice dataset for automatic speaker verification, derived from the Mozilla Common Voice corpus (Fig. \ref{fig:diagrams}). This dataset and the corresponding evaluation conditions address previous limitations of multilinguality, speaker diversity, data in the read-speech style, and public accessibility. 
The dataset comprises over 212,000 monolingual speakers across 81 languages and around 4,500 multilingual speakers across 40 unique languages. 

From the monolingual speakers, we derive the Tidy-M condition that covers all 81 languages with target (same-speaker) and non-target (different-speaker) trials matched in language. 
From the multilingual speakers, we derive the Tidy-X condition that covers 40 languages with target and non-target trials in both intra- and cross-lingual configurations.
To the best of our knowledge, this is the largest publicly available human-annotated dataset of its kind, offering broad speaker and language diversity (Table~\ref{tab:dataset_comparison}). 

In addition, we evaluate a series of ResNet-34 and ResNet-293 models, pre-trained and fine-tuned on combinations of VoxBlink2, VoxCeleb and TidyVoice to demonstrate the value of the proposed corpora. This approach yields a substantial reduction in Equal Error Rate (EER) across multiple benchmarks.
Finally, the datasets, splits, trial pairs, and trained models are publicly released to promote reproducible research and advance the development of multilingual speaker recognition and anti-spoofing systems.

\begin{table}[t]
\centering
\caption{Comparison of prominent multilingual datasets for speaker recognition. For each dataset, we report the number of speakers, languages, and utterances, along with the duration in hours, and domain.}
\label{tab:dataset_comparison}
\resizebox{0.48\textwidth}{!}{%
\begin{tabular}{l|l|l|l|l|l}
\hline
\textbf{Dataset} & \textbf{\# Spkr} & \textbf{\# Lang} & \textbf{\# Utt} & \textbf{Dur} & \textbf{Domain} \\
\hline
\hline
VoxCeleb~\cite{nagrani2017voxceleb} & 1,251 & Multi & 153,516 & 352 & YouTube \\
VoxCeleb2~\cite{chung2018voxceleb2} & 6,112 & Multi & 1,092,009 & 2,442 & YouTube \\
VoxBlink1~\cite{lin2024voxblink1} & 38,065 & 13 & 1,455,190 & 2,135 & YouTube \\
VoxBlink2~\cite{lin2024voxblink2} & 111,284 & 18 & 9,904,382 & 16,672 & YouTube \\
NIST SRE Superset~\cite{sadjadi2021nist} & 6,867 & 52 & 605,761 & 6,193 & CTS, AfV \\
\hline
\textbf{Our Work (Tidy-M)} & \textbf{212,617} & \textbf{81} & \textbf{5,616,611} & \textbf{8,138} & \textbf{Read}  \\
\quad Tidy-M: Train & 141,623 & 81 & 5.4M & 7,800 & Read  \\
\quad Tidy-M: Test & 70,994 & 81 & 218K & 350 & Read  \\
\textbf{Our Work (Tidy-X)} & \textbf{4,474} & \textbf{40} & \textbf{321,711} & \textbf{457} & \textbf{Read}  \\
\quad Tidy-X: Train & 3,666 & 40 & 262K & 370 & Read  \\
\quad Tidy-X: Test & 808 & 40 & 60K & 87 & Read  \\
\hline
\end{tabular}%
}
\end{table}

\section{Datasets and Conditions}
\label{sec:benchmarks}

The Mozilla Common Voice corpus (MCV) is a publicly available, massively multilingual, crowdsourced collection of read speech. 
While MCV provides a client ID as a proxy for a unique speaker identity, this identifier is not always reliable. 
As identified in our prior works ~\cite{hintz2024commonbench, zhang2025quantifying}, the client ID in MCV can suffer from speaker heterogeneity, where multiple individuals contribute recordings under a single ID. 

A verification-based pipeline was developed in \cite{zhang2025quantifying} to quantify and mitigate this issue for monolingual data across 76 languages, enabling the creation of cleaner speaker profiles. 
Specifically, a ResNet-293 speaker verification model pre-trained on VoxBlink2 and VoxCeleb2 was used to determine the cosine similarity between a designated enrollment file and all remaining test files for each client ID within a language. 
Files with a score less than 0.4 were assumed to come from different speakers and were thus excluded from the final dataset \cite{zhang2025quantifying}. 

In this work, we applied this method to an additional five languages for a total of 81 languages and assessed speaker heterogeneity among multilingual client IDs, that is, client IDs which appear in more than one language in MCV. 
We fully removed any ID where a significant number of these cross-lingual pairs produced cosine similarity scores lower than 0.2, indicating a high likelihood of a speaker identity switch. 
Through this process, we found that 433 out of 4,907 multilingual speakers (approximately 9\%) likely corresponded to different speakers across each language.


The Tidy-M partition of the TidyVoice dataset contains data from 81 languages from speakers who contributed data in a single language (referred to as ``monolingual'' speakers). 
In prioritizing a large training set, we assigned speakers with more than four utterances to the training set, while the remaining speakers were used for the test set, with the additional constraint that each language have at least five speakers in the test set. 
For the Tidy-M condition, target trials consisted of utterance pairs from the same speaker in the same language. Non-target trials consisted of utterance pairs from different speakers in the same language. The test set contains approximately 2.8 million pairs.

The Tidy-X partition of the TidyVoice dataset contains data from speakers who contributed data in two or more languages (referred to as ``multilingual'' speakers) from 40 different languages. We created the train and test splits in such a way that all languages were represented in both portions (Table \ref{tab:dataset_comparison}).
The Tidy-X benchmark trials include all four combinations of speaker and language pairings. 
Target trials contain utterance pairs produced by the same speaker in the same language and in different languages (same language: 2M, different language: 2M). 
Non-target trials contain utterance pairs from different speakers, again in the same language and in different languages (same language: 4M, different language: 4M).

\begin{table*}[h]
\centering
\caption{Results of developed models on datasets across different speech styles and audio sources. Each entry reports the performance in terms of EER (\%) / minDCF. 
Within the training data, VB2 corresponds to the VoxBlink2 pre-trained models and the plus (+) corresponds to additional fine-tuning with the TidyVoice partitions. Light gray corresponds to out-of-domain trials, while dark gray represents the proposed trial lists.}
\label{tab:overall_results_extended}
\resizebox{0.7\textwidth}{!}{
\begin{tabular}{@{}llcccccc@{}}
\toprule
\textbf{Model} & 
\textbf{Training data} & 
\textbf{Vox1-O} & \textbf{Vox1-E} & \textbf{Vox1-H} &
\textbf{\cellcolor{gray!30}CANDOR} & 
\textbf{\cellcolor{lightgray}Tidy-M} & 
\textbf{\cellcolor{lightgray}Tidy-X} \\
\midrule
\multirow[c]{4}{*}{ResNet-34} 
 & Tidy-M & 11.14 / 0.89 & 11.96 / 0.89 & 15.7 / 0.91 & \cellcolor{gray!30}4.86 / 0.32 & \cellcolor{lightgray}\textbf{0.39 / 0.06} & \cellcolor{lightgray}2.83 / 0.71 \\
 & VB2 \cite{lin2024voxblink2} & \textbf{0.39 / 0.02} & \textbf{0.62 / 0.07} & \textbf{1.11 / 0.10} & \cellcolor{gray!30}2.81 / 0.11 & \cellcolor{lightgray}3.34 / 0.41 & \cellcolor{lightgray}4.10 / 0.87 \\
 & \quad+ Tidy-M      & 1.15 / 0.12 & 1.29 / 0.14 & 2.39 / 0.22 & \cellcolor{gray!30}\textbf{2.00 / 0.10} & \cellcolor{lightgray}0.67 / 0.09 & \cellcolor{lightgray}\textbf{1.90 / 0.69} \\
 & \quad+ Tidy-X  & 3.37 / 0.35 & 3.86 / 0.36 & 6.52 / 0.50 & \cellcolor{gray!30}5.15 / 0.34 & \cellcolor{lightgray}1.85 / 0.21 & \cellcolor{lightgray}3.07 / 0.86  \\
\midrule
\multirow[c]{3}{*}{ResNet-293} 
 & VB2 \cite{lin2024voxblink2} & \textbf{0.30 / 0.02} & \textbf{0.57 / 0.05}      & \textbf{1.02 / 0.09}      & \cellcolor{gray!30}2.97 / 0.13 & \cellcolor{lightgray}2.74 / 0.82 & \cellcolor{lightgray}3.89 / 0.93 \\
 & \quad+ Tidy-M      & 0.53 / 0.06 & 0.83 / 0.08      & 1.54 / 0.14      & \cellcolor{gray!30}\textbf{1.60 / 0.07} & \cellcolor{lightgray}\textbf{0.35 / 0.19} & \cellcolor{lightgray}\textbf{1.65 / 0.71} \\
 & \quad+ Tidy-X  & 4.00 / 0.39 & 4.50 / 0.41      & 7.79 / 0.55      & \cellcolor{gray!30}5.53 / 0.39 & \cellcolor{lightgray}2.50 / 0.29 & \cellcolor{lightgray}3.72 / 0.38 \\
\bottomrule
\end{tabular}
}
\end{table*}

\section{Experimental setup}
\label{sec:exp_setup}

All experiments were conducted using the WeSpeaker open-source toolkit~\cite{wang2023wespeaker}, for which we describe the acoustic features, model architectures, training strategy, and evaluation protocol.

For all experiments, the input audio was first converted into 80-dimensional log Mel-filterbank energies. These features were extracted using a frame length of 25 ms and a hop size of 10 ms. During training, we used fixed-length segments of 200 frames. Utterance-level cepstral mean and variance normalization was applied to normalize the features.

We used two variants of the ResNet architecture~\cite{he2016deep} as our speaker embedding extractors consist of ResNet-34 and ResNet-293. The initial convolutional layer was followed by a series of residual blocks. The temporal resolution was progressively down-sampled by the convolutional stride, while the channel dimension was increased.
The final convolutional layer was followed by an Attentive Statistics Pooling~\cite{okabe2018attentive} layer, which aggregates frame-level features into a fixed-size utterance-level representation. A final fully-connected layer then projects this representation into a 256-dimensional speaker embedding.



The evaluated models were the ResNet-34 model trained from scratch on the Tidy-M partition, along with a series of models using the ResNet-34 and ResNet-293 architectures, developed in parallel from the pre-trained models \cite{lin2024voxblink2} with data from over 117,000 speakers from the VoxCeleb2~\cite{chung2018voxceleb2} and VoxBlink2~\cite{lin2024voxblink2} datasets.
This pre-trained model was then further fine-tuned on the Tidy-M and Tidy-X partitions of TidyVoice. 


Both the from-scratch Tidy-M and pre-trained VB2 models were trained using the Additive Angular Margin (AAM) loss, also known as ArcFace~\cite{deng2019arcface}. The training was performed for 150 epochs using the Adam optimizer with an initial learning rate of $10^{-3}$, which was reduced by a factor of 10 every 30 epochs. 

The +Tidy-M and +Tidy-X models were fine-tuned on the corresponding training partitions of the TidyVoice dataset. 
This stage adapted the generalized models to the specific characteristics and linguistic diversity of the TidyVoice data. The fine-tuning process also used the AAM loss, with the $\sim$141,600 speakers from Tidy-M or the $\sim$3,700 speakers from Tidy-X. We used a smaller initial learning rate to ensure stable convergence and decayed it similarly to the pre-training stage. 

For all training stages, we employed on-the-fly data augmentation to improve model robustness. This included adding background noise from the MUSAN corpus~\cite{snyder2015musan} and reverberation effects from the RIRS\_NOISES dataset~\cite{ko2017study}.

For evaluation, 256-dimensional embeddings were extracted for each utterance in the trial lists. The verification score for each pair was computed using cosine similarity. We report performance using two standard metrics, EER and Minimum Detection Cost Function (minDCF). EER refers to the point on the DET curve where the false acceptance rate (FAR) and false rejection rate (FRR) are equal. MinDCF is calculated with a target prior of $P_{target}=0.01$ and costs $C_{miss}=1$ and $C_{false}=1$.

Each model was evaluated on the three standard VoxCeleb1 benchmarks (Vox1-O, Vox1-E, Vox1-H)~\cite{nagrani2017voxceleb}, the recently designed CANDOR benchmark~\cite{reece2023candor, farhadipour2025adaptive}, which serves as a challenging out-of-domain test set. This benchmark consists of 93,000 trial pairs from approximately 190 speakers engaged in conversational, spontaneous speech. The speakers met online and held free discussions in English on a variety of topics. 


\section{Results and Discussion}
\label{sec:results}

The main results for all seven model variations across the six evaluation conditions are presented in Table~\ref{tab:overall_results_extended}.

We first evaluate the proposed models on the existing VoxCeleb benchmarks.
For these, the pre-trained models performed best. This outcome is not entirely surprising, as the pre-trained models were trained on VoxBlink2 and finetuned on VoxCeleb2 data, making them perfectly matched in style. However, specializing the models on our read-speech data introduced a domain mismatch, which likely resulted in a slight performance degradation on this specific ``in-the-wild'' task.

The models fine-tuned on our large Tidy-M partition exhibited a remarkable improvement on the CANDOR benchmark, which also consists of conversational and spontaneous speech. This is a critical finding, as it demonstrates that fine-tuning on the massive speaker diversity of Tidy-M facilitates generalization to the unseen, conversational domain.

This pattern of results holds for both the ResNet-34 and ResNet-293 models, with the larger ResNet-293 architecture consistently outperforming the ResNet-34 architecture in almost all cases.

\begin{figure}[t]
  \centering
  \includegraphics[width=\linewidth]{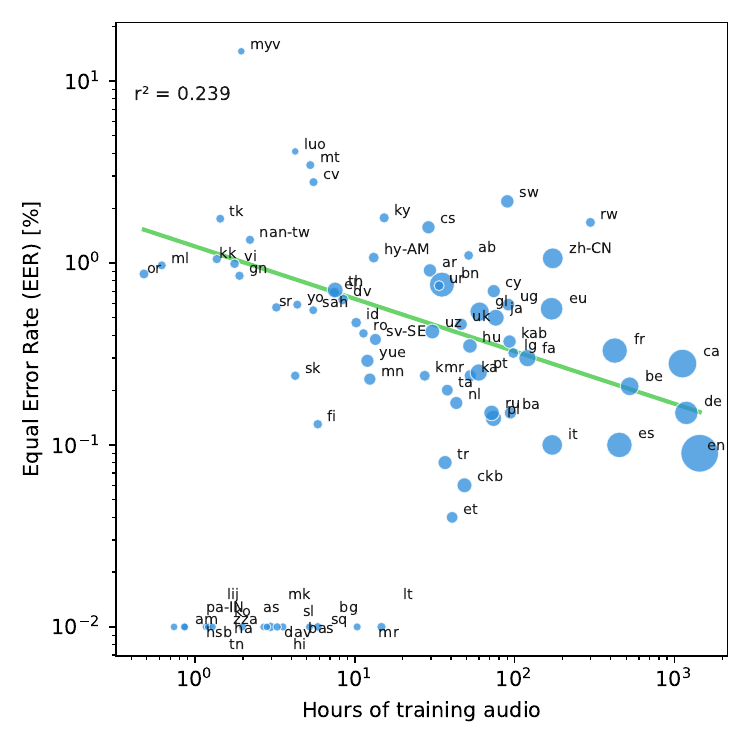}
  \caption{EER as a function of training audio hours per language (log–log scale). Point size reflects the number of test speakers, and labels denote language codes. The green line shows a power-law fit, excluding languages with zero EER. 
  }
  \label{fig:eer-vs-hours}
\end{figure}

For the Tidy-M condition, the best performance involves models trained on the Tidy-M training partition. Among ResNet-34 models, this was the Tidy-M model trained from scratch, followed closely by the fine-tuned model. Among ResNet-293 models, the fine-tuned +Tidy-M model had the best performance. 
The language-specific results from this condition are visualized in Figure~\ref{fig:eer-vs-hours}. While several languages showed near-perfect performance after fine-tuning, these results should be interpreted with caution, as some languages had a very small number of evaluation trial pairs.

The true impact of our fine-tuning process was also evident in the remarkable performance gains for languages where the pre-trained VB2 model struggled. For instance, the EER for Odia (``or'') improved from 29.51\% to 0.87\%, a relative improvement of 97\%. Similarly, the EER for Czech (``cs'') improved from 26.67\% to 1.57\%, and for Kabyle (``kab'') from 15.43\% to 0.37\%.

This widespread enhancement strongly supports our central hypothesis: fine-tuning on the massive speaker diversity of the Tidy-M partition forces the model to learn more fundamental and robust speaker-specific characteristics that transcend superficial linguistic cues. Rather than simply overfitting to read speech, the model becomes a more powerful generalist, better able to handle unseen conversational domains and a wide array of languages. This is a direct testament to the value of our large-scale benchmark.

Finally, the Tidy-X condition was designed to test the model's ability to handle cross-lingual speaker verification, with a simultaneous comparison to intra-lingual speaker verification. The fine-tuned +Tidy-M model performed best on this condition. The improved performance relative to the fine-tuned +Tidy-X model was likely due to the substantially larger size of the Tidy-M training dataset. Nevertheless, the fine-tuned +Tidy-X model still offered numerical improvement on this condition compared to the baseline model. 

In addition, the Tidy-X condition offered insight into how exactly the models handled the distinctions between target and non-target distributions from the same and different languages. In this trial list, the hardest case was distinguishing target speaker–different language from non-target–same language, while the easiest was target–same language versus non-target–different language.

The few exceptions mainly occurred when fine-tuning on the smaller Tidy-X, where we suspect that the higher capacity of the ResNet-293 made it more prone to overfitting on the limited speaker set, shifting the task from speaker discrimination toward language effect cancellation, compared to the more constrained ResNet-34. In addition, the fine-tuned +Tidy-M model had a clear advantage in terms of the number of speakers, while the fine-tuned +Tidy-X model provided a theoretically interesting case where the model specifically learned from a smaller set of multilingual speakers. 

\section{Conclusion}
\label{sec:conclusion}

In this paper, we introduced the TidyVoice dataset for multilingual automatic speaker verification from a tidied and curated version of the Mozilla Common Voice corpus. 
Our experiments yielded several key findings. First, fine-tuning on our large-scale Tidy-M partition provided a performance boost compared to the baseline model on the TidyVoice intra-lingual (Tidy-M) and mixed language (Tidy-X) conditions. 
Second, and critically, we demonstrated that this fine-tuning process enhanced model generalization: despite being trained on read speech, our model outperformed its pre-trained baseline on a test set derived from the out-of-domain, conversational CANDOR corpus. 

To facilitate further progress, we are publicly releasing the full TidyVoice dataset and evaluation trials, as well as all developed models \footnote{https://github.com/areffarhadi/wespeaker/tree/master/examples/tidyvocie}. 
Given the steady growth of MCV, we foresee the possibility to dynamically expand this dataset over the long-run with straightforward extensions to additional languages and speakers. 

\bibliographystyle{IEEEbib}
\bibliography{strings,refs}

@inproceedings{desplanques20_interspeech,
  title     = {{ECAPA-TDNN}: Emphasized Channel Attention, Propagation and Aggregation in {TDNN} Based Speaker Verification},
  author    = {Brecht Desplanques and Jenthe Thienpondt and Kris Demuynck},
  year      = {2020},
  booktitle = {Interspeech 2020},
  pages     = {3830--3834},
  doi       = {10.21437/Interspeech.2020-2650},
  issn      = {2958-1796},
}

@inproceedings{chung2018voxceleb2,
  title={{VoxCeleb2}: Deep speaker recognition},
  author={Chung, Joon Son and Nagrani, Arsha and Zisserman, Andrew},
  booktitle={Proc. Interspeech},
  year={2018}
}

@inproceedings{nagrani2017voxceleb,
  title={Vox{C}eleb: a large-scale speaker identification dataset},
  author={Nagrani, Arsha and Chung, Joon Son and Zisserman, Andrew},
  booktitle={Interspeech 2017},
  year={2017}
}

@article{far2024analys,
  title={Analysis of deep generative model impact on feature extraction and dimension reduction for short utterance text-independent speaker verification},
  author={Farhadipour, Aref and Veisi, Hadi},
  journal={Circuits, Systems, and Signal Processing},
  volume={43},
  number={7},
  pages={4547--4564},
  year={2024},
  publisher={Springer}
}

@inproceedings{lin2024voxblink1,
  title={Vox{B}link: A large scale speaker verification dataset on camera},
  author={Lin, Yuke and Qin, Xiaoyi and Zhao, Guoqing and Cheng, Ming and Jiang, Ning and Wu, Haiying and Li, Ming},
  booktitle={ICASSP 2024-2024 IEEE International Conference on Acoustics, Speech and Signal Processing (ICASSP)},
  pages={10271--10275},
  year={2024},
  organization={IEEE}
}

@inproceedings{lin2024voxblink2,
  title     = {{VoxBlink2}: A 100K+ Speaker Recognition Corpus and the Open-Set Speaker-Identification Benchmark},
  author    = {Yuke Lin and Ming Cheng and Fulin Zhang and Yingying Gao and Shilei Zhang and Ming Li},
  year      = {2024},
  booktitle = {{Interspeech 2024}},
  pages     = {4263--4267},
  doi       = {10.21437/Interspeech.2024-1490},
  issn      = {2958-1796},
}

@article{farhadipour2025adaptive,
  title={Adaptive Multimodal Person Recognition: A Robust Framework for Handling Missing Modalities},
  author={Farhadipour, Aref and Vukovic, Teodora and Dellwo, Volker and Motlicek, Petr and Madikeri, Srikanth},
  journal={arXiv preprint arXiv:2512.14961},
  year={2025}
}

@inproceedings{hutiri2022bias,
  title={Bias in automated speaker recognition},
  author={Hutiri, Wiebke Toussaint and Ding, Aaron Yi},
  booktitle={Proceedings of the 2022 ACM conference on fairness, accountability, and transparency},
  pages={230--247},
  year={2022}
}

@article{reece2023candor,
  title={The CANDOR corpus: Insights from a large multimodal dataset of naturalistic conversation},
  author={Reece, Andrew and Cooney, Gus and Bull, Peter and Chung, Christine and Dawson, Bryn and Fitzpatrick, Casey and Glazer, Tamara and Knox, Dean and Liebscher, Alex and Marin, Sebastian},
  journal={Science Advances},
  volume={9},
  number={13},
  pages={eadf3197},
  year={2023},
  publisher={American Association for the Advancement of Science}
}

@article{sadjadi2021nist,
  title={{NIST SRE CTS S}uperset: A large-scale dataset for telephony speaker recognition},
  author={Sadjadi, Seyed Omid},
  journal={arXiv preprint arXiv:2108.07118},
  year={2021}
}

@inproceedings{he2016deep,
  title={Deep residual learning for image recognition},
  author={He, Kaiming and Zhang, Xiangyu and Ren, Shaoqing and Sun, Jian},
  booktitle={Proceedings of the IEEE conference on computer vision and pattern recognition},
  pages={770--778},
  year={2016}
}

@inproceedings{okabe2018attentive,
  title={Attentive statistics pooling for deep speaker embedding},
  author={Okabe, Koji and Koshinaka, Takafumi and Shinoda, Koichi},
  booktitle={Proc. Interspeech 2018},
  pages={2252--2256},
  year={2018}
}

@inproceedings{deng2019arcface,
  title={Arcface: Additive angular margin loss for deep face recognition},
  author={Deng, Jiankang and Guo, Jia and Xue, Niannan and Zafeiriou, Stefanos},
  booktitle={Proceedings of the IEEE/CVF conference on computer vision and pattern recognition},
  pages={4690--4699},
  year={2019}
}

@article{snyder2015musan,
  title={{MUSAN}: A music, speech, and noise corpus},
  author={Snyder, David and Chen, Guoguo and Povey, Daniel},
  journal={arXiv preprint arXiv:1510.08484},
  year={2015}
}

@inproceedings{ko2017study,
  title={A study on data augmentation of reverberant speech for robust speech recognition},
  author={Ko, Tom and Peddinti, Vijay and Povey, Daniel and Khudanpur, Sanjeev},
  booktitle={2017 IEEE International Conference on Acoustics, Speech and Signal Processing (ICASSP)},
  pages={5220--5224},
  year={2017},
  organization={IEEE}
}

@inproceedings{zhang2025quantifying,
  title     = {Quantifying and Reducing Speaker Heterogeneity within the {Common Voice} Corpus for Phonetic Analysis},
  author    = {Miao Zhang and Aref Farhadipour and Annie Baker and Jiachen Ma and Bogdan Pricop and Eleanor Chodroff},
  year      = {2025},
  booktitle = {{Interspeech 2025}},
  pages     = {3933--3937},
  doi       = {10.21437/Interspeech.2025-2027},
  issn      = {2958-1796},
}

@inproceedings{hintz2024commonbench,
  title={{CommonBench}: A larger Scale Speaker Verification Benchmark},
  author={Hintz, Jan and Siegert, Ingo},
  booktitle={Proc. SPSC 2024},
  pages={17--20},
  year={2024}
}

@inproceedings{wang2023wespeaker,
  title={{WeSpeaker}: A research and production oriented speaker embedding learning toolkit},
  author={Wang, Hongji and Liang, Chengdong and Wang, Shuai and Chen, Zhengyang and Zhang, Binbin and Xiang, Xu and Deng, Yanlei and Qian, Yanmin},
  booktitle={ICASSP 2023-2023 IEEE International Conference on Acoustics, Speech and Signal Processing (ICASSP)},
  pages={1--5},
  year={2023},
  organization={IEEE}
}

@article{barahona2025analysis,
  title={Analysis of {ABC} Frontend Audio Systems for the {NIST-SRE24}},
  author={Barahona, Sara and Silnova, Anna and Mo{\v{s}}ner, Ladislav and Peng, Junyi and Plchot, Old{\v{r}}ich and Rohdin, Johan and Zhang, Lin and Han, Jiangyu and Palka, Petr and Landini, Federico and others},
  journal={arXiv preprint arXiv:2505.15320},
  year={2025}
}

@article{lee20232022,
  title={The 2022 {NIST} language recognition evaluation},
  author={Lee, Yooyoung and Greenberg, Craig and Godard, Eliot and Butt, Asad A and Singer, Elliot and Nguyen, Trang and Mason, Lisa and Reynolds, Douglas},
  journal={arXiv preprint arXiv:2302.14624},
  year={2023}
}

@article{nagrani2020voxceleb,
  title={{VoxCeleb}: Large-scale speaker verification in the wild},
  author={Nagrani, Arsha and Chung, Joon Son and Xie, Weidi and Zisserman, Andrew},
  journal={Computer Speech \& Language},
  volume={60},
  pages={101027},
  year={2020},
  publisher={Elsevier}
}

\end{document}